# Fabrication technology for high light-extraction ultraviolet thin-film flip-chip (UV TFFC) LEDs grown on SiC

Burhan K. SaifAddin[1*], Abdullah Almogbel[1,3], Christian J. Zollner[1], Humberto Foronda[1], Ahmed Alyamani[3], Abdulrahman Albadri[3], Michael Iza[1], Shuji Nakamura[1,2], Steven P. DenBaars[1,2,] and James S. Speck[1]

[1]Materials Department, UCSB, CA 93106, USA
[2]Department of Electrical and Computer Engineering, UCSB, CA 93106, USA
[3]King Abdulaziz City of Science and Technology (KACST), Saudi Arabia
*bks@ucsb.edu

**Abstract**
The light output of deep ultraviolet (UV-C) AlGaN light-emitting diodes (LEDs) is limited due to their poor light extraction efficiency (LEE). To improve the LEE of AlGaN LEDs, we developed a fabrication technology to process AlGaN LEDs grown on SiC into thin-film flip-chip LEDs (TFFC LEDs) with high LEE. This process transfers the AlGaN LED epi onto a new substrate by wafer-to-wafer bonding, and by removing the absorbing SiC substrate with a highly selective $SF_6$ plasma etch that stops at the AlN buffer layer. We optimized the inductively coupled plasma (ICP) $SF_6$ etch parameters to develop a substrate-removal process with high reliability and precise epitaxial control, without creating micromasking defects or degrading the health of the plasma etching system. The SiC etch rate by $SF_6$ plasma was ~46 µm/hr at a high RF bias (400 W), and ~7 µm/hr at a low RF bias (49 W) with very high etch selectivity between SiC and AlN. The high $SF_6$ etch selectivity between SiC and AlN was essential for removing the SiC substrate and exposing a pristine, smooth AlN surface. We demonstrated the epi-transfer process by fabricating high light extraction TFFC LEDs from AlGaN LEDs grown on SiC. To further enhance the light extraction, the exposed N-face AlN was anisotropically etched in dilute KOH. The LEE of the AlGaN LED improved by ~3X after KOH roughening at room temperature. This AlGaN TFFC LED process establishes a viable path to high external quantum efficiency (EQE) and power conversion efficiency (PCE) UV-C LEDs.

## 1. Introduction

AlGaN ultraviolet light-emitting diodes (UV LEDs) and ultraviolet laser diodes (UV LDs) in the range of 265–280 nm are needed to develop novel disinfection and sterilizing technologies (water, air, and surfaces) to improve access to clean water [1–4], improve public health [5,6], and enable other biotech applications [7–10]. UV LEDs are a viable technology for replacing mercury gas discharge lamps in disinfection and biotechnology applications [11–18]. Hospitals, for example, reduced the rates of hospital-acquired drug-resistant infections by 25% using ultraviolet disinfection (via mercury lamps (254 nm)), as reported by various clinical trials [19,20]. Furthermore, the AlN or AlGaN epi-transfer and heterogeneous integration technology can be employed to enhance the performance of photodiodes (PDs) [21], high electron mobility transistors (HEMTs) [22,23], bulk acoustic resonators (BARs) [24–26], and high-aspect-ratio SiC microstructures and devices [27–30].

Researchers developing AlGaN LEDs have made significant progress in the last 30 years [11,18,31–44]; however, the technology is limited by light extraction efficiency (LEE) [45–54]. The LEE of AlGaN LEDs is less than that of InGaN blue LEDs for five main reasons. First, the absence of a transparent current spreading layer over p-GaN, such as indium tin oxide (ITO) for blue LEDs, requires the use of flip-chip (FC) LEDs architecture with a highly-reflective p-side down. Second, the poor conductivity of Mg-doped AlGaN necessitates the use of an optically absorbing $p^+$-GaN layer for hole injection into the Mg-doped AlGaN and the formation of an ohmic p-contact; GaN has an absorption coefficient of $1.5 \times 10^5$ cm$^{-1}$ at 275 nm [55]. Third, the reflectivity of ohmic p-contacts, is limited to ~70-80% , in the 265-280 nm range. Fourth, the light from the AlGaN MQW emitters at these wavelengths is ~40% transverse magnetic (TM) polarized [47,48,56–58], which propagates laterally; finite-difference time-domain (FDTD) calculations estimate that the extraction of TM emission is >10X less efficient than transverse electric (TE) polarization in typical volumetric FC LEDs [48] (in volumetric FC LEDs light has to travel through the transparent growth substrate) [48]. Lastly, the encapsulants used to enhance LEE in blue LEDs suffer low transparency, poor stability, and a low refractive index in deep ultraviolet LEDs [59–63].

The most commonly employed substrates for AlGaN UV-C LEDs are sapphire and AlN but they yield LEDs with low LEE. Although sapphire substrates are transparent, their thermal conductivity is limited, and they lack efficient ways to extract TM polarization because roughening the sidewalls of sapphire LEDs dies is challenging [46]. AlN substrates grown by physical vapor transport (PVT) are limited by an absorption around 264 nm (4.7 eV) with an absorption coefficient of about ~35 cm$^{-1}$ [64] whereas hydride vapor phase epitaxy (HVPE) grown AlN substrates have less absorption (~10 cm$^{-1}$) [65]. Researchers have largely overlooked AlGaN LEDs grown on SiC substrates [66–74] because SiC absorbs strongly below its optical bandgap (3.2 eV and 3.0 eV for 4H-SiC and 6H-SiC, respectively), but this disadvantage can be overcome with a novel thin-film flip-chip (TFFC) LED architecture in which the SiC is removed with a highly selective SF$_6$ chemical plasma etch between SiC and AlN.

Growing AlN on SiC substrates is promising due to their similar crystallographic structure, polarity, chemical stability, and low lattice mismatch (0.8%) [75–80]. AlN grown on SiC with low threading dislocations has been demonstrated with MOCVD [78,81] and plasma-assisted MBE [71,76]. Furthermore, we demonstrate in this paper that AlGaN LEDs grown on SiC can be processed into thin-film LEDs with high LEE [82–85]; for example, FDTD simulations by Ryu et al. show that TM emission's LEE in textured thin-film LEDs is significantly higher (>6X) than in volumetric AlGaN FC-LEDs [48]. InGaN thin-film blue LEDs were developed with very high LEE using laser lift-off for substrate removal, N-face GaN KOH photoelectrochemical (PEC) roughening, and a p-side reflective mirror [86–88]. Several methods for laser lift-off of AlGaN UV-C LEDs have been developed [89–94]; however, laser-induced melting of the Al in the dislocated interface of the AlN buffer layer in AlGaN LEDs causes more cracks than laser lift-off of the GaN buffer layer in InGaN LEDs [95] because the Al melts at 660 °C, whereas Ga melts at 30 °C [91]. Employing PEC etching to lift-off thin-film LEDs from their growth substrates [70,71] is more difficult with AlN than with GaN because N-face AlN etches anisotropically in KOH (without above-bandgap light assistance) considerably faster (~20x) than N-face GaN does [98]. Thus, KOH roughening can damage the active region before the lateral lift-off etch is completed.

In this paper, we demonstrate a process to fabricate high extraction efficiency TFFC LEDs grown on SiC. We characterize SiC and AlN etching using inductively coupled plasma (ICP) $SF_6$ plasma. The $SF_6$ gas flow, process pressure, and sample RF bias power were investigated to determine first-order trends between the etch parameters and the etch rates of SiC and AlN. Next, the etch parameters were optimized to develop a highly reliable SiC substrate-removal process with no micromasking defects and precise epitaxial control, without degrading the plasma etching system health. A highly selective $SF_6$ etch between SiC and AlN (90:1) was achieved which was essential for removing the SiC substrate and exposing a smooth AlN surface (roughness ~0.6 nm). We demonstrate high LEE thin-film flip chip (TFFC) AlGaN LEDs grown on SiC. We compare the power output of TFFC AlGaN LEDs before and after KOH roughening of the exposed AlN surface. The AlGaN LED's LEE improved by ~3X after roughening in KOH [0.25 M] for 70 sec at room temperature.

## 2. SiC substrate thinning characteristics

AlN was grown on quartered 2-inch 6H-SiC substrates (SiCrystal AG) by MOCVD as described elsewhere [81]. A one-step lapping process was used to thin the SiC substrates after bonding them to a carrier with high thermal conductivity (4.9 W/cm a-plane, 3.9 W/cm c-plane) with n- and p- bonding pads using Au/Au thermocompression bonding [73,74]. The SiC based samples were mounted episide-down on a 2-inch stainless-steel chuck using a wax (that melts at 120 °C), which provides mechanical adhesion during lapping. The lapping process was performed using South Bay Technologies multipurpose lapping system (Model 920). The SiC growth substrates were thinned from 250 µm to 75 µm by lapping with a 9 µm Dia-Grid Diamond Disc (Allied High Tech Products, Inc.). The water-cooled lapping tool had a lapping rate of 29 µm/hr using 9 µm grit diamond discs. A summary of the thinning parameters is shown in Table 1. The lapping damage generated by lapping with a 9 µm grit was approximately 3 x 9 µm = 27 µm, which avoids damage to the active region. The mechanical lapping yields a total thickness variation (TTV) of about 15–20 µm across a quartered 2-inch wafer.

## 3. $SF_6$ etch chemistry characteristics

The SiC substrates were plasma etched in a Panasonic ICP etching system (E6261) using $SF_6$ plasma [67,68,101–103]. ICP etching systems have independent ICP and RF bias sources. The ICP plasma (generated by a current flowing in a planner coil above the etch chamber) power controls the ions and radicals densities. The RF bias power controls the ions bombardment energy and thus is proportional to sputter (physical) etching. The samples were mounted episide-down on a 6-inch carrier wafer using a vacuum diffusion pump fluid, which provides thermal conduction between the sample and the carrier wafer. The sample surface temperature depends on the lower electrode temperature and on the plasma ions' energies and densities. The lower electrode of the ICP system was backside cooled by pressurized He, and the carrier wafer temperature was kept at 11 °C during all etches. The 6-inch carrier wafer was a 1.5 mm thick fused silica wafer with a 100 nm Al backside coating to hold the fused silica carrier wafer to the electrostatic chuck. The chamber and carrier wafer were cleaned in $O_2$ plasma and then seasoned with a 2 min $SF_6$ etch at 1000 W ICP power and low RF bias (49 W). We found that chamber and carrier wafer seasoning suppressed the formation of SiC pillars. The SiC etch rates were determined by profilometry (Dektak) and scanning electron microscopy (SEM).

Etching SiC with $SF_6$ plasma produces the volatile products $SiF_4$, $SiF_2$, $CF_2$ and $CF_4$ [104]; the absence of etch-induced polymer generation renders the long etching (for several hours, if needed) consistent and reliable. Thus, long $SF_6$ etching does not need to be interrupted to clean the etch chamber because the etch chemistry does not cause polymer accumulation. However, the carrier wafer's temperature in the ICP etch chamber should be controlled during long etches because elevated temperatures can increase the SiC etch rate.

We developed a two step $SF_6$ ICP etch process to remove SiC: (1) high SiC etch rate process; (2) high selectivity SiC:AlN etch process. The two-step etch parameters are summarized in Table 2.

To achieve the highest etch rates of bulk SiC, we observed SiC etch rate as a function of $SF_6$ flow and process pressure. Figure 1(a) shows that the SiC etch rate increased as $SF_6$ flow increased, with improved etch uniformity. Figure 1(b) shows that the etch rate was sensitive to the process pressure. Increasing the process pressure, increases the ions' densities but decreases the ions' energies, by decreasing its mean free path. The SiC etch rate data in figure 2(b) shows that a chamber pressure of 1.33 Pa resulted in the highest SiC etch rate at the following process parameters: 1000 W ICP power, 400 W RF bias power, and 50 sccm $SF_6$ flow.

To achieve high etch selectivity, we utilized the fact that SiC etches chemically in $SF_6$, whereas AlN does not etch chemically in $SF_6$ and is only etched at a low rate by $SF_6$ sputter etching (at low RF bias). Figure 2(a) shows that the SiC etch rate increased as the RF bias power increased. At high RF bias power, $SF_6$ sputter etch dominates whereas at low RF bias (below 60 W), $SF_6$ chemical etch dominates and the sputter etch rate for SiC and AlN is very low. The etch rate at 400 W was about ~40 µm/hr when measured over 40 min and 46 µm/hr when measured over 2 hr. The SiC substrate was etched first at ~46 µm/hr over 1.2 hr to remove about 60 µm of SiC; the temperature of the He-cooled carrier wafer increased slightly from 11 °C to 15 °C. Achieving higher SiC etch rate is possible in plasma etch systems that have higher ICP power and RF bias power.

Figure 2(b) shows that the SiC:AlN etch selectivity was very sensitive to RF bias, especially below 60 W. . At high RF bias, the etch selectivity was low (10:1 at 400 W) because at this regime, sputter etching has low selectivity between SiC and AlN. On the other hand, the SiC:AlN etch selectivity increased significantly as the RF bias decreased below 60 W, where the $SF_6$ chemically etches SiC but does not etch AlN due to the formation of the low-volatility species $AlF_3$, which was demonstrated previously in studies of selective reactive ion etching of GaAs on AlGaAs and GaP on AlGaP in fluorine containing plasma [105–108]. However, the strong dependence of the SiC:AlN etch selectivity on the RF bias power decreased when the RF bias power was less than 47 W, as the SiC etch rate decreased more than the AlN etch rate did.

SiC:AlN etch selectivity is essential for removing the SiC substrate without damaging the active layer by stopping at the AlN buffer layer with precise epitaxial control. The etch selectivity was dependent on the substrate temperature (maintained at 11 °C), process pressure, and the RF bias power, especially in the 47–49 W region. For a thinned, quartered 2-inch SiC substrate with ~15–20 µm in TTV, a SiC:AlN etch selectivity of 90:1 at 49 W RF bias and 1.33 Pa was sufficient to reliably and selectively etch SiC and expose a pristine surface of N-face AlN. At a lower process pressure of 0.8 Pa, the SiC:AlN etch selectivity increased to 150:1. This indicates that the selectivity can be increased further by optimizing the process pressure. Senesky et al. [27] reported an $SF_6$ etch selectivity (at a low RF bias) of 16:1 (SiC:AlN) using AlN deposited with reactive sputtering; the higher bulk SiC etch selectivities obtained with MOCVD-grown AlN (at a low RF bias) indicate that AlN grown by MOCVD could replace Ni as a hard mask in fabricating high-aspect-ratio SiC microstructures and devices [28,29].

We optimized the $SF_6$ ICP etch to ensure that it removed the thick SiC substrate without producing micromasking defects on the thin-film LEDs surface and without degrading the plasma etching system health. First, we avoided the use of commonly used metal carriers inside the etch chamber (refer to Table 3). In ICP plasma etch systems, $SF_6$ etching is typically performed using a 6-inch Ni or Al carrier wafer because: both have low etch rates in $SF_6$ [101,109], as shown in Table 3. However, the use of metal carrier wafers increases micro-masking defects (which produces 40+ µm SiC pillars, as shown in figure 3(a)) that will not etch completely, even with the selective etch. Furthermore, sputtered metal from metal carriers can affect subsequent etches or cause electrical shorting [27], and require manual cleaning of the etch chamber after every etch. The use of fused silica ($SiO_2$) or sapphire carrier wafers minimized the micro-masking defects (SiC pillars) on the etched surface and did not affect the ICP etch chamber walls. The suppression of SiC pillars is shown in SEM images in figure 3(b). Selective etching of a SiC surface with suppressed micromasking defects exposed a smooth AlN surface (refer to figure 4(e) for an AFM scan of the exposed AlN surface with RMS roughness < 1 nm) that was free of SiC pillars, as shown in the optical and SEM micrographs of the TFFC LEDs in figure 5. Some literature indicates that the etch pressure requires optimization to remove SiC pillars; however, we found that pillar formation can be suppressed with appropriate selection of the carrier wafer, combined with the seasoning of both the carrier wafer and the etch chamber.

## 4. TFFC UV-C LED (278 nm) demonstration

The thin-film transfer technology was demonstrated on a high LEE TFFC UV-C LED, as shown in figures 4-6.

Figure 4 shows the process flow to fabricate TFFC LEDs from LED grown on SiC. The AlGaN LED structure was first grown on a quartered 2-inch 6H-SiC substrate (SiCrystal AG) by MOCVD as reported elsewhere [85]. The LED structure is shown in figure 6(a) and consisted of AlN (3.2 µm), $Al_{0.8}Ga_{0.2}N$ (180 nm), n-$Al_{0.6}Ga_{0.4}N$ (1.1 µm), 278 nm MQWs 4x($Al_{0.39}Ga_{0.61}N$/ $Al_{0.6}Ga_{0.4}N$), p-$Al_{0.53}Ga_{0.47}N$ (50 nm), and p-GaN (5 nm). The samples were cleaved into 1.1 x 1.1 $cm^2$ samples and were processed into LEDs. The n-contact was V/Al/V/Au (20/80/20/200 nm); the p-contact was unannealed Ni/Al/Ni/Au (1/150/100/1000 nm) (refer to [85] for details). After a 3 min solvent clean and 30 sec $O_2$ plasma clean (100 W), the LEDs were bonded to a thermally conductive substrate (semi-insulating 4H-SiC, Cree, Inc.) with n- and p- bonding pads (20/1500 nm Ti/Au were deposited by e-beam). The samples were bonded using low-temperature Au/Au thermocompression bonding. A Finetech flip-chip bonder (FINEPLACER lambda) was employed for aligned wafer-to-wafer bonding. The bond was performed at 30 N/$cm^2$ for 5 min at 275 °C in air. Then, the samples were bonded further in custom-designed graphite fixtures that applied ~300 N/$cm^2$ pressure while the sample was annealed at 200 °C for 2 hr in air (the pressure was maintained during a 5 min cooldown); refer to Table 4 for a summary of the low-temperature Au/Au thermocompression bonding parameters. The 6H-SiC growth substrate was lapped mechanically to 75 µm and removed by a two-step $SF_6$ plasma etch, as previously described. Figure 5(a) shows plain-view optical micrograph images of processed UV LEDs with partially exposed AlN during an etch interrupt. The exposed TFFC LEDs had no visible cracks or chipping. The complete removal of SiC and exposure of pristine AlN surface was visible to the eye — a colorful interference pattern was observed as the light incident, and reflected from different surfaces in the TFFC LED interferes [110]. Also, the wettability of the surface changed as the N-face AlN was exposed. Namely, the surface of the etched carbon-face SiC was hydrophilic, and the surface of the exposed AlN was hydrophobic, as shown in figure 5(a). The TFFC LED contacts design is shown in figure 5(b) which shows a 5-finger topology for p-contact, surrounded by n-contact, with p-contact area of 0.093 $mm^2$. The 5-finger p-contact topology was adopted to avoid current crowding in the n-AlGaN layer [111], which had a relatively high resistivity (~60 mΩ-$cm^2$).

After complete removal of the SiC growth substrate, the suspended AlN/AlGaN/n-AlGaN thin-film in-between the TFFC LEDs, which is above the n- and p-pads (as shown in figure 5(b)), can be removed via a patterned hard mask ($SiO_2$) and KOH, or by etching a wider and deeper mesa around the LEDs mesa (into 80% of the AlN thickness) — before FC bonding. However, we relied on the residual tensile stress in the AlN/AlGaN/n-AlGaN film, which caused it to slightly concave upward [98]. Thus, after the TFFC LEDs were singulated, the n- and p- pads becomes accessible to wire bonding, as shown in figure 5(c). The TFFC LEDs were singulated by mechanical sawing using an ADT 7100 Dicing Saw. The TFFC LEDs were covered by photoresist for protection from the cooling/cleaning water jets that are needed during mechanical sawing. Subsequently, the TFFC LEDs were mounted on TO headers using a Ni-Au epoxy from Dexerials Corp., and the light output of the LEDs was measured in a 75 mm integrating sphere (Digital Instruments, Inc.).

Figure 6(b) compares the dependence a TFFC LED's light output power (L) on the injected DC current (I) before and after optimized KOH roughening of the exposed AlN surface [85]. The slope of the power versus current increased by ~3X after KOH roughening of the exposed AlN layer in dilute KOH (0.25 M) for 70 sec (refer to the bird's-eye-view SEM image of the resulting hexagonal pyramids in figure 6(c)). The LEE enhancement after KOH roughening was limited to ~3X (as estimated from the enhancement in the L-I slope) due to the limited p-contact reflectivity and the use of a 5 nm $p^+$-GaN layer (a strongly absorbing layer) to inject holes into the LED active region. Further LEE enhancement is possible without using p-GaN and by using a more reflective p-contacts. For example, the photoluminescence (PL) of MQW wells in an n-i-n structure (n-Al$_{0.6}$Ga$_{0.4}$N/286-nm-MQWs/n-Al$_{0.6}$Ga$_{0.4}$N/Al$_{0.6}$GaN/AlN) could be enhanced by ~3.9X [112] after KOH roughening by using a highly reflective p-contact (a Pt/Al/Ni/Au p-contact with a Pt thickness of 0.26 nm and reflectivity of 90% at 286 nm was used to produce a ~3.9X intensity enhancement in PL after KOH roughening).

The LEE of TM emissions in KOH-roughened TFFC LEDs are expected to be higher than in bulk FC LEDs. For example, Lee et al. studies on LEE in AlGaN UV LEDs [113] showed that LEE from a PSS sapphire AlGaN LED die was limited by sidewall roughness, more so than in InGaN-based bulk FC LEDs, and that LEE from sidewalls is significant in thick (e.g., 300 μm thick sapphire) bulk FC AlGaN LED but becomes negligible in relatively thin bulk FC AlGaN LEDs (e.g., 90 μm thick sapphire). However, in TFFC LEDs, the contribution of sidewall emissions is negligible because the thin-film LED widths are much larger than its thickness, which is only 2–3 μm, and both TE and TM light emission couple into the roughened AlN surface; for example, FDTD simulations by Ryu et al. showed that the LEE of TM emission in roughened AlGaN thin-film LEDs is significantly higher (>6X) than in bulk AlGaN FC-LEDs [48]. The TE/TM emissions ratio depends on the MQWs strain and the Al composition in the wells and barriers [114–116]. We speculate that the strain state of the MQWs grown on SiC will be similar to MQWs grown on sapphire because (1) the AlN buffer layer is normally relaxed in both sapphire and SiC substrates, and therefore the AlGaN MQWs are similarly compressively strained by the AlN lattice; and (2) the piezoelectric component due to thermal coefficient of expansion (TCE) mismatch (compressive in sapphire, and slightly tensile in SiC) between the AlN and AlGaN layers is negated because it affects them similarly.

Estimating the LEE in unroughened thin-film LEDs could be challenging. For example, thin-film blue LEDs with KOH-roughened GaN buffer layers, ray tracing simulations estimate 13 % LEE for each single-pass extraction (6 bounces for full extraction) [117], whereas wave optics simulations estimate 31% LEE for each single-pass extraction (3 bounces for full extraction) [118]. We will discuss our estimates for the LEE of TFFC LEDs in a future publication.

After the optimized KOH roughening, the LED CW power was 7.8 mW at 95 mA, emitting at 278.5 nm, which yielded an EQE of 2%. Stable encapsulation [61] can further enhance the LEE from TFFC LEDs; however, most resins absorb strongly below 330 nm and decompose over time, which renders them commercially unviable. A higher LEE for AlGaN TFFC LEDs is also achievable if the p-GaN layer is replaced with a transparent hole injector [119–121] or if the reflectivity of the p- and n-metal contacts is increased [48,117,122,123]. The IV characteristics do not change with KOH etching as discussed in details elsewhere [98] (refer to Figure 6c); however, in the device discussed here the voltage increased by ~ 1V after 50 sec KOH etching (refer to Figure 6d) which was probably due to a crack in the TFFC LED; however, further KOH etching did not affect the IV.

## 5. Conclusions

We developed a highly selective $SF_6$ plasma etch of SiC over AlN and a viable manufacturing method for epitaxial transfer from SiC to another substrate via wafer bonding and SiC substrate-removal by $SF_6$ etching. Then, we demonstrated it on UV-C LEDs that were grown on SiC to fabricate high LEE TFFC LEDs. The SiC substrate was bonded via Au-Au thermocompression bonding to another n- and p-patterned thermally conductive substrate. The growth substrate was removed by a two-step ICP etching without micromasking defects on the etched surface. The first step was a high SiC etch rate (~46 μm/hr) at a high RF bias power (400 W). The second step was a selective SiC etch rate (~7 μm/hr) at a low RF bias power ( 49 W), with an etch selectivity of SiC:AlN ~90:1 at a process pressure of 1.33 Pa; higher SiC:AlN etch selectivity of ~150:1 was also achieved at a lower process pressure. The highly selective $SF_6$ etch at low bias was essential for removing all SiC by reliably stopping at the LEDs' AlN buffer layer, which acts as an etch stop layer. We demonstrated a TFFC manufacturing method for UV-C LEDs that were grown on SiC. The LEE was significantly enhanced via KOH roughening of the exposed N-face AlN. KOH roughening enhanced the LEE by ~3X for UV LEDs without encapsulation and despite the use of 5 nm p-GaN as a p-contact layer.

## Acknowledgements

This work was funded by the King Abdulaziz City for Science and Technology (KACST), the Technology Innovations Center (TIC) program, and the KACST-KAUST-UCSB Solid State Lighting Program. The authors are appreciative of the support of the Solid State Lighting and Energy Electronics Center (SSLEEC) at UCSB. A portion of this work was conducted in the UCSB nanofabrication facility and part of the NSF NNIN network (ECS-0335765), as well as the UCSB MRL, which is supported by the NSF MRSEC Program (DMR05-20415). The authors acknowledge the UCSB-Collaborative Research in Engineering, Science and Technology (CREST) Malaysia project. This work was also supported by the National Science Foundation Graduate Research Fellowship Program (Grant No. 1650114). Any opinions, findings, and conclusions or recommendations expressed in this material are those of the author(s) and may not reflect the views of the National Science Foundation. The authors would also like to thank the cleanroom staff at UCSB nanofabrication facility, especially Brian Thibault for helpful discussions.

## List of Figures

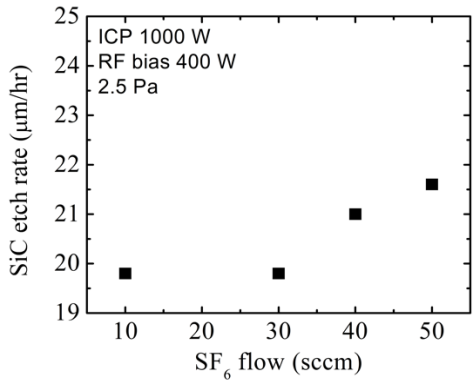

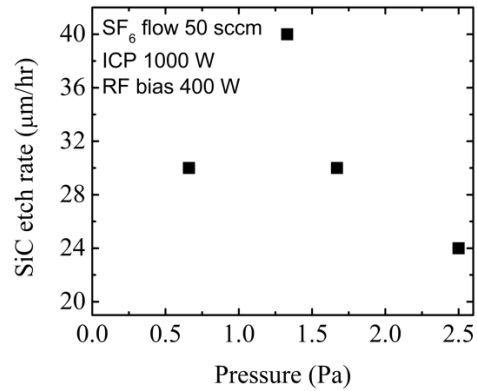

(a) Bulk SiC etch rate (μm/hr) vs SF6 flow rate (sccm). Higher flow rate increases the etch rate and improves the lateral etch uniformity.

(b) Bulk SiC etch rate (μm/hr) vs pressure (Pa). The etch rate peaks at 1.33 Pa when the product of the ions' energies and the ion densities is maximum.

**Figure 1.** The $SF_6$ flow and process pressure, were examined to determine the trends between the etch parameters to optimize for a high SiC etch rate. The process parameters were fixed at 1000 W ICP, 400 RF bias, and $SF_6$ 50 sccm flow. The etch rates were measured for bulk SiC over a period of ~40 min – on a fused silica carrier wafer.

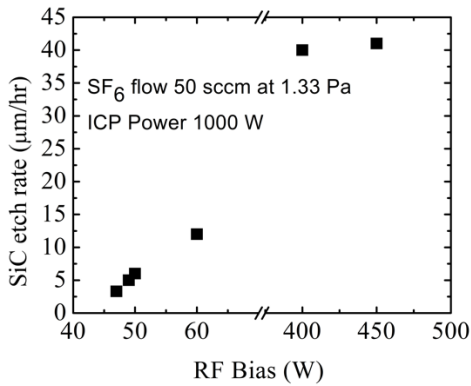

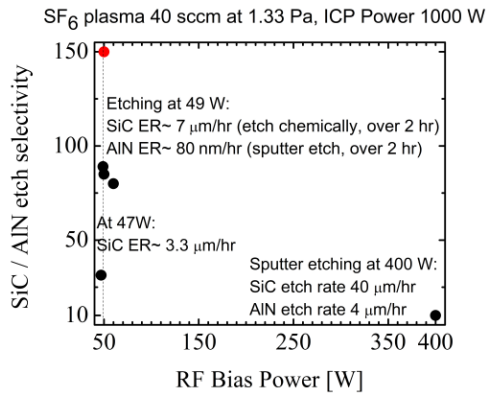

(a) Bulk SiC etch rate (μm/hr) vs RF bias (W).

(b) SiC:AlN etch selectivity as function of RF bias power (W).

**Figure 2.** In figure 2(a), at high RF bias, the SiC etch rate was dominated by sputter etching, however, at low RF bias sputter etch rate is minimal, and SiC is primarily chemically etched. Figure 2(b) shows that SiC:AlN etch selectivity has a strong dependence on RF bias below 60 W. The SiC:AlN etch selectivity at 49 W RT bias was 90:1 during a 2 hr etch, however, the selectivity decreased below 47 W. A higher SiC:AlN etch selectivity 150:1 (shown in red) was measured at a lower process pressure (0.8 Pa).

**Table 1.** Summary of lapping parameters used to mechanically thin SiC substrate with 9 µm diamond grid disc.

| Lapping settings | Lapping etch rate |
|---|---|
| CCW<br>5 wafer rotation/ min<br>6 pad rotation /min | 25-30 µm/min |

**Table 2.** Summary of etch parameters for the two-steps ICP $SF_6$ etch to selectively remove SiC (using fused silica as a carrier).

| Process parameter | Fast SiC etch | Selective SiC etch (slow) |
|---|---|---|
| Pressure | 1.33 Pa | 1.33 Pa |
| ICP power | 1000 W | 1000 W |
| $SF_6$ Flow | 50 sccm | 50 sccm |
| ICP bias | 400 W | 49 W |
| Etch rate | 40 µm/hr (40 min)<br>46 µm/hr (90 min)<br>47 µm/hr (2 hr) | ~7 µm/hr<br>(2 hr) |
| SiC:AlN Selectivity | 10:1 | ~90:1<br>(~150:1 at 0.8 Pa) |
| Impact on SiC surface | Rough hydrophilic SiC surface | After SiC is completely etched, the surface changes from hydrophilic to hydrophobic; smooth N-face AlN surface (RMS roughness < 1 nm) is exposed. |

**Table 3.** Summary of the etch rate, and selectivity of SiC and various 6-inch carrier wafers used in the ICP system (at 1000 W ICP, 400 W RF bias, 1.33 Pa).

| 6" Carrier wafer | Etch rate (µm/hr) | SiC:Carrier wafer etch selectivity at 400 W RF bias | Cost | Comments |
|---|---|---|---|---|
| SiC | 40 | 1:1 | $1000 | -- |
| Sapphire | 5 | 8:1 | $400 | Satisfactory selectivity but expensive. |
| Fused Silica | 20 | 1.9:1 | $30 | Satisfactory selectivity and inexpensive. |
| Al wafer | 1.5-2 | 25:1 | $200 | Sputter into sample and etch chamber walls |
| Ni wafer | 1 | 40:1 | $200 | Sputter into sample and etch chamber walls. |
| Cu film on Si wafer | NA | NA | NA | Sputter contaminating and nonvolatile etch byproducts into sample and etch chamber walls |
| Ni film on Si wafer | 1 | 40:1 | NA | Sputter into sample and etch chamber walls. Also, Ni films thicker than 2 µm buckle and delaminate due to high compressive stress. |

**Table 4.** Summary for Au-Au thermo-compression bonding characteristics with Finetech flip-chip bonder.

| Bonding type | Temperature and Force | Sample size |
|---|---|---|
| Au-Au thermo-compression bonding | To minimize p-mirror damage: 1-275 ºC, 5 min (30 N/cm²) 2- 200 ºC, 2 hr (300 N/cm² with graphite fixture) | 0.3x0.3 cm² to 1.5x1.5 cm² |

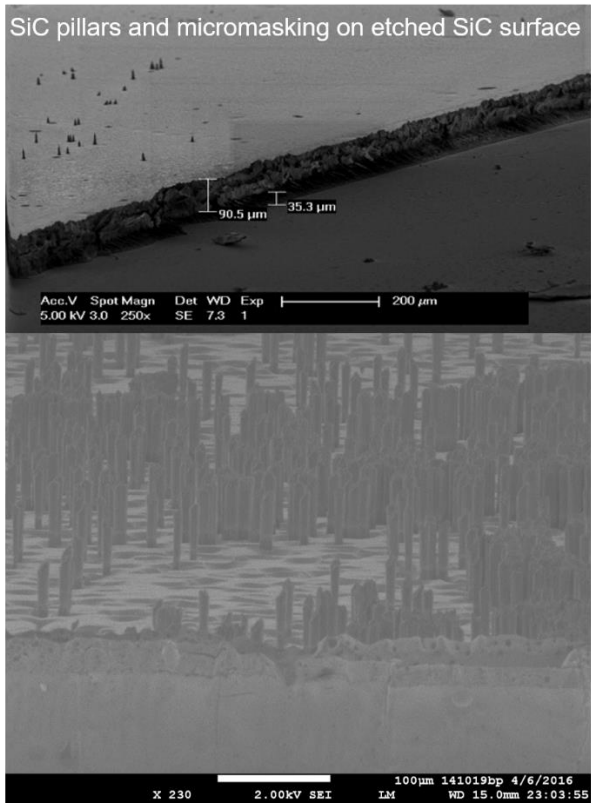

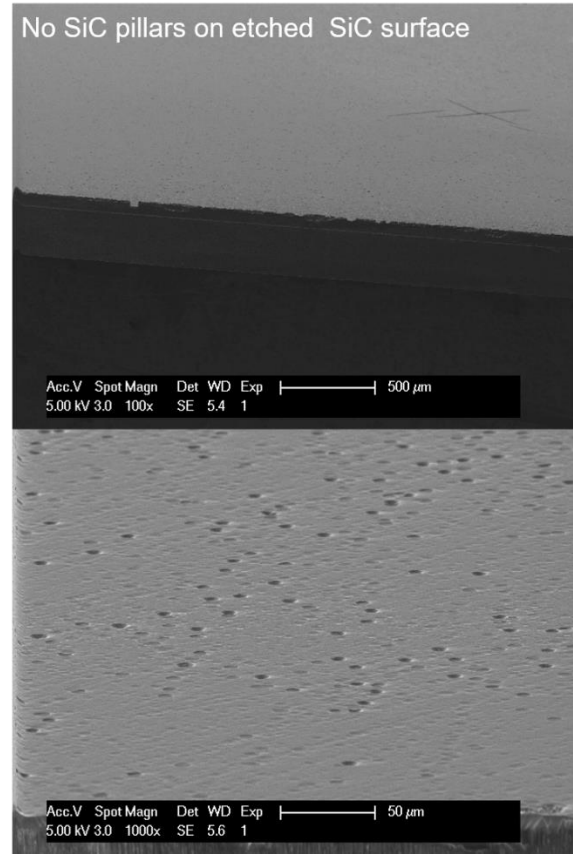

(a)            (b)

**Figure 3.** SiC pillars can be completely removed from the surface if appropriate process controls are applied. The image on the left (a) shows SiC pillars (40+ µm mircomasking defects) that can form during etching and that will not be completely etched even when the highly selective etch is subsequently applied. The SiC pillars formation was suppressed as shown on the right (b) by an optimized etch and process controls: 1) Metal carrier wafers were avoided (fused silica carrier was employed). 2) The carrier wafer and etch chamber were seasoned with SF$_6$ plasma prior to etching.

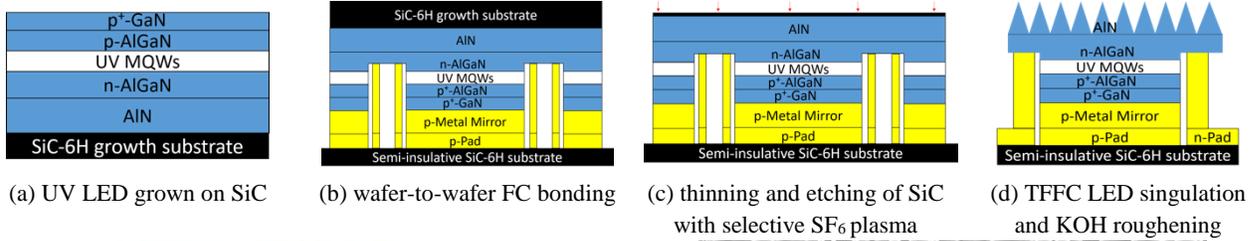

(a) UV LED grown on SiC

(b) wafer-to-wafer FC bonding

(c) thinning and etching of SiC with selective SF$_6$ plasma

(d) TFFC LED singulation and KOH roughening

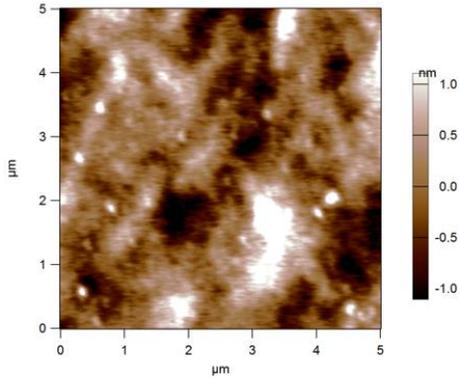

(e) AFM image of the surface of exposed AlN surface as (roughness ~0.6 nm) after complete etching of SiC in step (c).

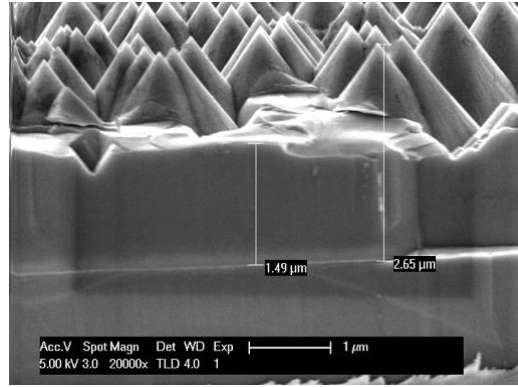

(f) Cross-sectional SEM image of nano-sharp hexagonal pyramids that are produced after KOH roughening of N-face AlN in step (d).

**Figure 4.** Process flow (a-d) demonstrate UV-C thin-film flip-chip (TFFC) grown on SiC and (e) shows AFM of exposed AlN surface after completing SiC growth substrate-removal in step (c). The hexagonal pyramids shown in (f) expand the effective angle of the light-extraction-cones. Figure adapted, with permission, from Ref. [85].

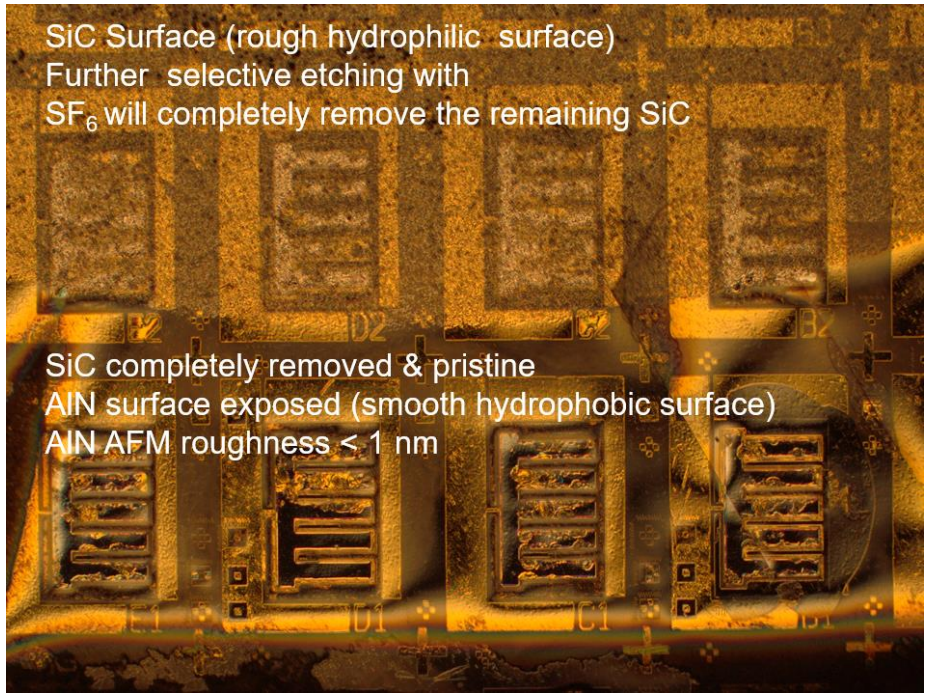

(a) Differential interference contrast (DIC) micrograph shows eight TFFC LEDs during an etch stop before the substrate is completely etched by the $SF_6$ plasma. In the top four LEDs, a thin layer of SiC remained. In the bottom four LEDs, no SiC remained and the N-face AlN was exposed. The colorful fringes in the lower four LEDs were due to thinfilm interference.

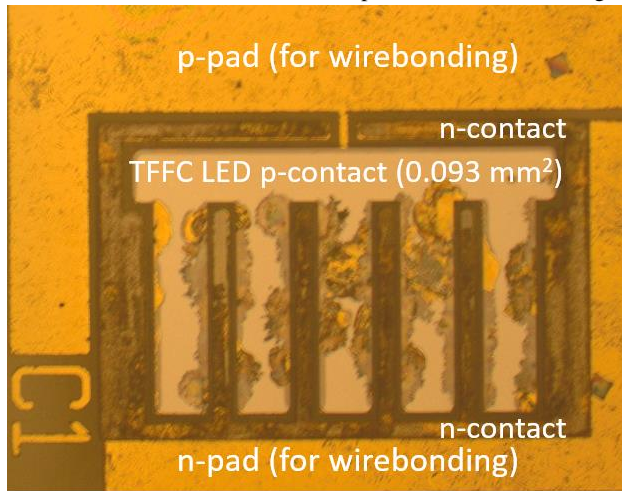

(b) Optical micrograph of UV-C TFFC LED (278 nm).

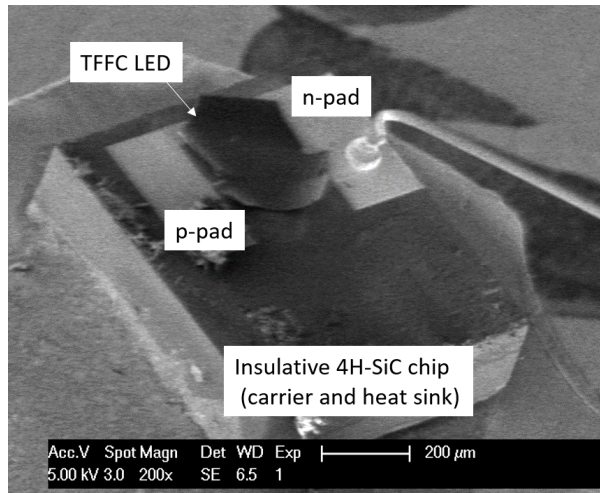

(c) SEM image of a packaged UV-C TFFC LED (278 nm).

**Figure 5.** Optical microscope micrographs of a processed thin film LED (5-finger topology) with no cracks or chipping in the active region in (a) and (b). After dicing, the suspended AlN/AlGaN/n-AlGaN between the TFFC LEDs concave up due to residual tensile stresses in the AlN/AlGaN/n-AlGaN film which renders the n- and p- pads accessible as shown in (c).

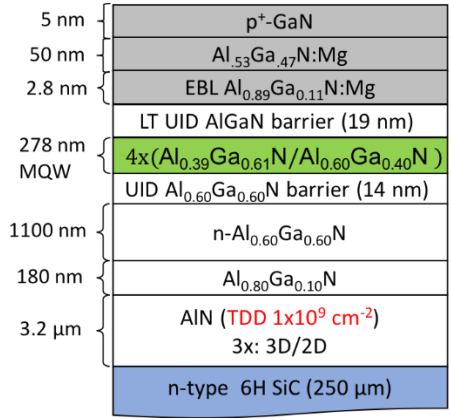

(a) 278 nm LED structure on SiC. The TDD was ~1x10$^9$ cm$^{-2}$ in the AlN buffer and AlGaN layers.

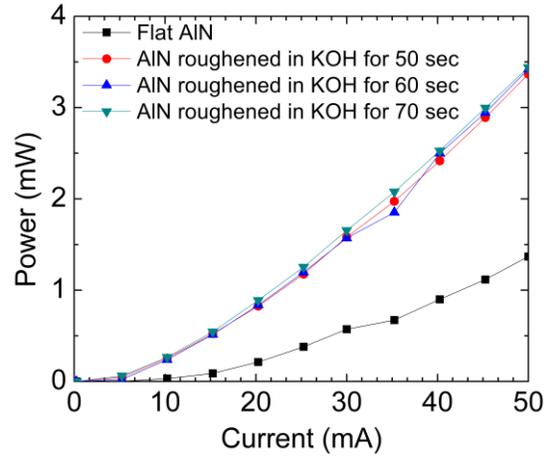

(b) The light-current (L-I) curve before, and after KOH roughening. The enhancement in the L-I slope KOH roughening was ~ 200%. The area of the p-contact (Ni/Al/Ni/Au) was 0.093 mm$^2$ ~0.1 mm$^2$.

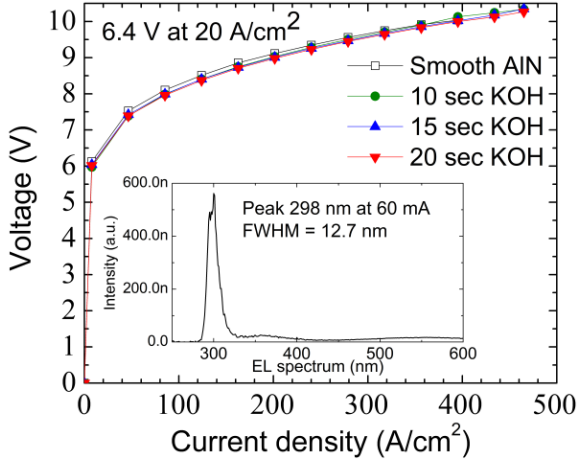

(c) The voltage-current (I-V) curve before, and after KOH roughening, for a 298 nm TFFC LED. The inset shows a 298 nm EL spectrum at 60 mA with FWHM of 12.7 nm. The area of the p-contact (Ni/Al/Ni/Au) was 0.013 mm$^2$.

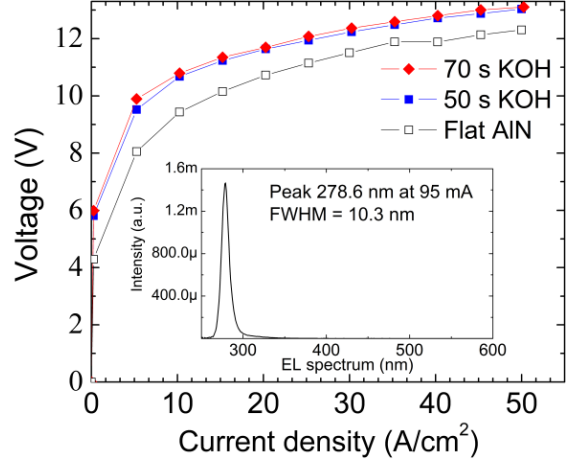

(d) The voltage-current (I-V) curve before, and after KOH roughening, for the 278 nm TFFC LED. The inset shows a 278.5 nm EL spectrum at 95 mA with FWHM of 10 nm. The DC power was 7.6 mW at 95 mA.

**Figure 6.** TFFC LEDs demonstration had high LEE. The LEE was enhanced significantly via KOH roughening of the exposed N-face AlN layer of TFFC LED. The L-I slope increased by ~3X with optimized KOH roughening without encapsulation, and despite using 5 nm p-GaN as p-contact layer.